# Amplification of high harmonics in 3D semiconductor waveguides

Dominik Franz[1], Rana Nicolas[1], Willem Boutu[1], Liping Shi[2], Quentin Ripault[1], Maria Kholodtsova[1], Bianca Iwan[1], Ugaitz Elu Etxano[3], Milutin Kovacev[2], Jens Biegert[3,4] and Hamed Merdji[1*]

[1] *LIDYL, CEA, CNRS, Université Paris-Saclay,*
*CEA Saclay 91191 Gif sur Yvette France*

[2] *Leibniz Universität Hannover, Institut für Quantenoptik,*
*Welfengarten 1, D-30167 Hannover, Germany*

[3] *ICFO – The Institute of Photonic Sciences,*
*Mediterranean Technology Park, Av. Carl Friedrich Gauss 3, 08860 Castelldefels, Spain*

[4]*ICREA, Pg. Lluís Companys 23, 08010 Barcelona, Spain*

[*]*correspondence to: hamed.merdji@cea.fr*

**Nanoscale amplification of non-linear processes in solid-state devices opens novel applications in nano-electronics, nano-medicine or high energy conversion for example. Coupling few nano-joules laser energy at a nanometer scale for strong field physics is demonstrated. We report enhancement of high harmonic generation in nano-structured semiconductors using nanoscale amplification of a mid-infrared laser in the sample rather than using large laser amplifier systems[1–4]. Field**

**amplification is achieved through light confinement in nano-structured semiconductor 3D waveguides. The high harmonic nano-converter consists of an array of zinc-oxide nanocones. They exhibit a large amplification volume, 6 orders of magnitude larger than previously reported and avoid melting observed in metallic plasmonic structures[2,3,5]. The amplification of high harmonics is observed by coupling only 5-10 nano-joules of a 3.2 μm high repetition-rate OPCPA laser at the entrance of each nanocone. Harmonic amplification (factor 30) depends on the laser energy input, wavelength and nanocone geometry.**

Ultrafast nano-photonics science is emerging thanks to the extraordinary progresses in nano-fabrication and ultrafast laser science. Boosting extremely intense electric fields in nano-structured photonic devices has the potential of creating nano-localized sources of energetic photons or particles opening vast applications in science and in the industry. Among those sources, high harmonic generation (HHG) is one of the most spectacular. This technology can not only revolutionize attosecond science but also prepare a new generation of optoelectronic devices. The high intensity needed for HHG in solids requires large laser amplification systems[6–10]. However, nanoscale field enhancement is nowadays a technology used to stimulate high-field phenomena[11,12]. Different meta-surfaces of metallic[13–15], dielectric and semiconductor[16–19] as well as metallic/dielectric and metallic/semiconductor hybrid structures[20–23] were successfully used to boost nonlinear processes. But attempts to amplify HHG in gases have led to controversial observations[1,4,5,24]. Due to the high density of atoms in solid crystals compared to gases, and the lower intensities of lasers needed to generate high harmonics[9,10,25,26], solids are promising candidates for efficient HHG assisted by field enhancement.

In 2016, Han et al. showed the amplification of high harmonics of an 800 nm femtosecond laser oscillator due to plasmonic enhancement in hybrid sapphire-metal waveguides[2]. More recently, amplification of high harmonics from localized enhancement induced by metallic nano-rods arranged at the surface of a thin film of silicon has been observed[3]. Even though plasmonic resonances ensure the highest enhancement of the local electric field, still, both works report major limitations. First, metallic nano-structures are easily damaged at high intensities; second, the amplification volume is extremely small and, finally, transmission through metal at the low wavelengths of the harmonics is very low which limits the non-linear conversion efficiency[2,3,27].

We propose a novel semiconductor 3D waveguide designed to efficiently couple and confine laser electric fields. Figure 1a shows an SEM picture of our zinc-oxide (ZnO) nanocone sample. In Figures 1b–d, FDTD simulations of the optimized waveguide geometry with a nanocone base of 4 µm are reported. The electric field confinement in the nanocones has been studied as a function of the laser wavelength (see supplementary movie M1). A homogeneous mode is found at our working laser wavelength of 3.2 µm. The light is strongly confined when reaching the middle of the nanocones which leads to significant enhancement. The nanocone exhibits a large amplification volume reaching about 1 µm$^3$ with a maximum intensity enhancement of 21. HHG amplification in the nanocones has been studied using an intense femtosecond 3.2 µm OPCPA system operating at 160 kHz repetition rate (see methods)[28]. A layout of the experimental setup is shown in Figure 2. The intensity regime is adjusted by attenuating the laser beam to pulse energies around a micro-joule in an 80 µm diameter focal spot, leading to peak intensities below a TW/cm$^2$. At this intensity, we are set below the ZnO crystal damage threshold. The harmonic emission can be imaged in the near-field (sample exit) and the far-field by a combination of two lenses and a CCD-camera. The interaction length along the laser propagation axis is about a micron size so that each nanocone acts as a point-like source. In the far field, the spatial profile is a coherent superposition of the individual nano-sources (see Figure S5). This gives an undisturbed view into the microscopic "single-atom" response. Near-field images of the 7$^{th}$ harmonic without and with nanocones at the same intensity of 0.35 TW/cm$^2$ are displayed in Figure 3a and 3b respectively. The harmonic source size emitted from the bare crystal (Figure 3a) is 40 µm at full width at half maximum (FWHM), which is comparable to the total size of the nanocones array (30 µm, Figure 3b). While the 7$^{th}$ harmonic from the bare crystal is weak, the signal emitted by the nano-emitters is intense. We note that the harmonic intensity varies from one nano-source to the other. This is due to the various shapes of the nanocones and to the fact that each nanocone is seeded by a varying local intensity in the laser focal spot. The supplementary movie M2 shows the amplification when the array of nanocones scans across the laser beam.

We have studied the amplification factor as a function of the laser intensity. The harmonic intensity profile along column A is shown in red in Figure 3c for low (0.35 TW/cm$^2$) from Figure 3a and 3b and high (0.80 TW/cm$^2$) pump intensity. At low pump intensity the harmonic signal from the nano-emitters is amplified by more than one order of magnitude (Figure 3c). Although the harmonic signal is still amplified at high pump intensity, the amplification factor is significantly smaller (Figure 3d). The intensity-dependent amplification factors of nanocones from columns A, B and C are displayed in Figure 4. Each column has a given amplification intensity dependence which is directly related to the sample fabrication process. Indeed, the nanocones milled at later times will deposit atoms on nanocones produced earlier (see supplementary information). The amount of re-deposition increases from column A (last patterned column) to column E (first patterned column) and will create a layer (see Figure S2) that will affect the light coupling and will scatter the emitted harmonics leading to a lower amplification. Thus, column A that has almost no re-deposition exhibits the highest amplification whereas column E shows almost no amplification. As a general trend, the amplification factor increases when the nanocones quality increases, as illustrated in Figure 4 from columns C to A, respectively.

In column A, at an intensity of 0.35 TW/cm$^2$, the local amplification of the 7$^{th}$ harmonic is high and reaches a maximum of 30 (strong amplification regime). Nanocone A4 exhibits the highest harmonic amplification factor. The amplification decreases with increasing pump intensity and saturates (weak amplification regime). Indeed, at high intensity, the harmonic emission saturates and a plateau is achieved[10]. The harmonic lateral source size of a single nano-emitter is estimated to be around 2 µm, i.e. below the driving laser wavelength of 3.2 µm which indicates the sub-wavelength confinement and amplification of the HHG emission. In the strong amplification regime, each nanocone couples between 5 nJ and 10 nJ which leads to a harmonic emission of about $2 \times 10^7$ photons/s per nanocone. This corresponds to a local conversion efficiency of $7 \times 10^{-9}$ for nanocone A4. The divergence of a single nano-source is estimated to be 0.5 rad.

Although the beam size ($5.10^3$ µm² at FWHM) was much larger than the coupling base size of the array of nanocones ($3.10^2$ µm²) and only a fraction of the beam was enhanced, the total signal emitted by the 25 nanocones was almost one order of magnitude larger in the strong amplification regime compared to the bare crystal. For future applications, the amplification of harmonics can be optimized for specific wavelengths. Movies M3 to M7 show amplification optimizations from an isolated nanocone with the coupling base ranging from 1.6 to 5 µm and wavelengths from 800 nm to 4 µm. While small nanocones do not couple the radiation efficiently, larger ones achieve optimum coupling with field enhancement up to 8 (64 in intensity). In all cases, shorter wavelengths propagate well towards the tip of the nanocone.

We have evaluated the damages induced by the high intensity pulses on the nanocones (see supplementary information). For intensities below a TW/cm² where resonant metallic nano-structures usually melt in less than few seconds[2,3,24], the harmonic signal emitted from the nanocones was stable during days. Above a TW/cm², a peeling of the re-deposited ZnO can be observed after a week of experiment (see Figure S4), but without affecting the measurements. Indeed, when increasing the intensity above 1 TW/cm² the amplification decreases. However, when lowering the intensity to few $10^{11}$/cm² the amplification rises up again to previously measured values even a week after. The robustness of the nanocones confirms the high potential for sustainable HHG applications.

In conclusion, we have shown that semiconductor nanocones have the potential to amplify HHG through local enhancement of the electric field by 3D waveguiding effects. We achieve 3D intense sub-wavelength size localization of amplified harmonics. This extends the demonstration of the 2D grating structures which showed amplification of the 5$^{th}$ harmonic emission[29]. In contrast to semiconductor nano-structures, metallic nano-structures damage quickly, and have much less enhancement volume and strongly absorb harmonic radiation[3]. Moreover, the control of the laser phase gradient should also favor tailoring the spatial and spectral phases of the harmonic emission, mandatory for the emission of attosecond pulses[30]. We also showed the dependence of the amplification on the fundamental laser intensity, where higher amplifications are measured for low intensities. The nanoscale amplification holds promise to replace external laser amplifiers using few nano-joules femtosecond laser system to boost non-linear processes in the strong field regime. It is then possible to scale this experiment to much higher repetition rate (up to GHz) lasers. While an array of nanocones is used for this demonstration, simulations show that a single nanocone could be used to generate an isolated HHG nano-source.

**Acknowledgments**

We acknowledge financial support from the European Union through LASERLAB ICFO002295 proposal (submitted on June 26[th] 2016 and accepted October 25[th] 2016 - https://laserlab.mbi-berlin.de/access/publish/listJointExperimentProjects.jsf). We acknowledge support from the VOXEL FET Open H2020 support, from the French ministry of research through the 2013 ANR grants "NanoImagine", 2014 "IPEX", 2016 "HELLIX" and from the C'NANO research program through the NanoscopiX grant, and the LABEX "PALM" through the grants "Plasmon-X" and "HILAC". We acknowledge the financial support from the French ASTRE program through the "NanoLight" grant. We acknowledge the financial support from the Swedish Research Council (Vetenskapsrådet) grant number 637-2013-439/D0043901 and the Swedish Foundation for International Cooperation in Research and Higher Education (STINT). Financial support by the Deutsche Forschungsgemeinschaft, grant KO 3798/4-11 and from Lower Saxony through "Quanten- und Nanometrologie" (QUANOMET), project NanoPhotonik are acknowledged. U.E. and J.B. acknowledge financial support from the Spanish Ministry of Economy and Competitiveness (MINECO), through the "Severo Ochoa" Programme for Centres of Excellence in R&D (SEV-2015- 0522) and the Fundació Cellex Barcelona.

We acknowledge discussions with Thierry Auguste. Finally, we acknowledge Franck Fortuna and Laurent Delbecq for access and support to the focused ion beam on the CSNSM laboratory (IN2P3, Paris Saclay University).

## Authors contribution

D.F. R.N., W.B., L.S., Q.R., M.Kh and H.M. carried out the experiment. The samples were produced by W.B. and Q.R.. The laser system and the facility were operated by U.E.E and J.B.. Simulations were performed by D.F., R.N., B.I. and Q.R.. Data analysis was performed by D.F., R.N., W.B., L.S., U.E.E. and H.M. H.M. proposed the physical concept. All authors discussed the results and contributed to the writing of the manuscript.

## Competing financial interest

The authors declare no competing financial interest.

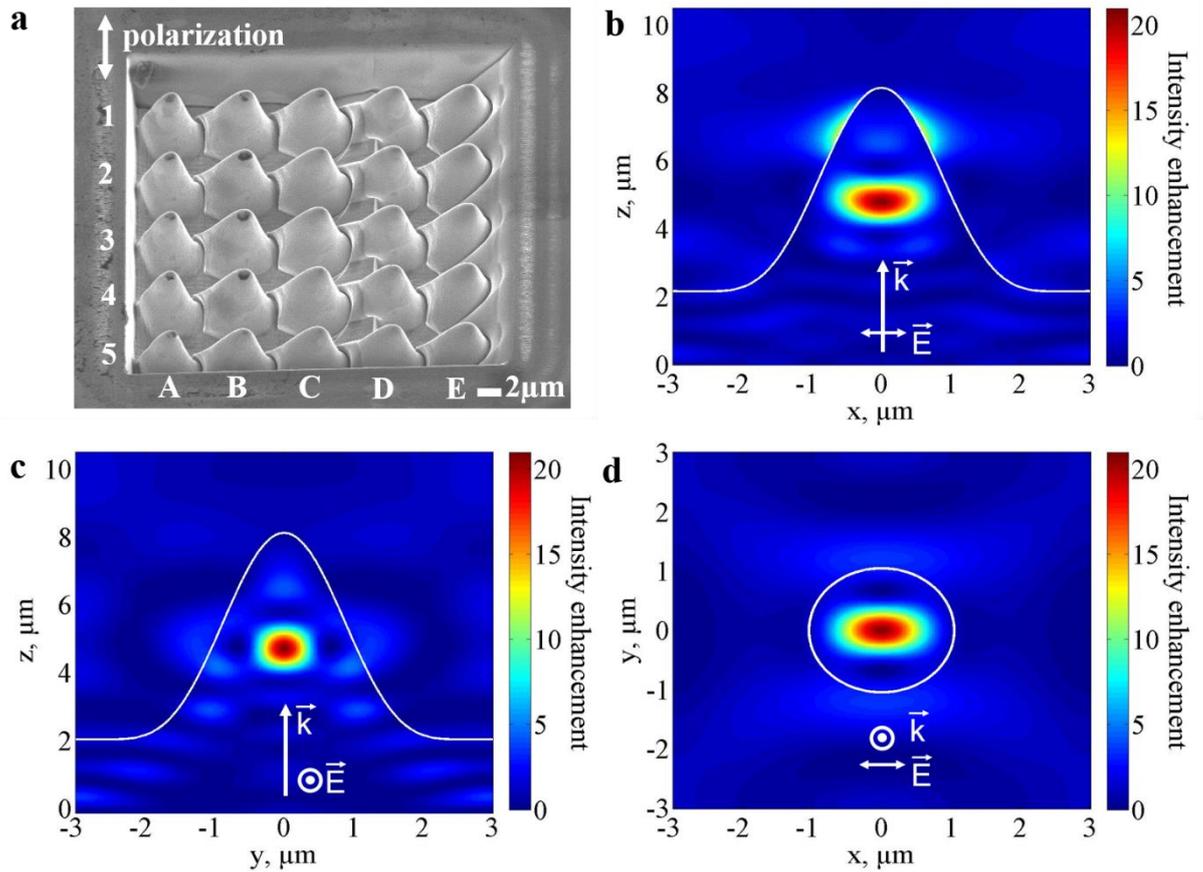

**Figure 1 | Sample properties and FDTD simulations. a,** SEM picture taken at an angle of 35° with respect to the crystal surface. Column A has the smallest dimension, with a base diameter of 4 µm whereas columns B to E have a larger shape (base varying from 4.5 to 5.5 µm). This is due to atomic re-deposition during patterning process with a Focused Ion Beam (FIB) (see supplementary information). All cones have a height of 6 µm. The image was taken after a month laser irradiation. The darker area at the tip of nanocones from columns A and B are due to the phase transition due to the long term irradiation (see supplementary information). The re-deposition of ZnO on early milled

columns (E to B respectively) has enlarged the size of the nanocones. Column E is strongly immersed by a thick layer of amorphous ZnO. The laser polarization follows the columns axis. **b, c, d,** Simulated distribution of the intensity enhancement in the x-z-plane, y-z-plane and x-y-plane, respectively, zoom on a single nanocone. The laser propagates along the z axis and is polarized along the x axis. White solid lines represent the borders of the nanocones. The nanocones are illuminated from the bottom at a wavelength of 3.2 µm. Beam propagation and polarization directions are indicated by k and E, respectively. The electric field couples into the nanocone and then propagates towards the tip. Due to the geometry of the nanocones, the angle of the incident light with respect to the nanocone surface exceeds the angle of total internal reflection (32°).

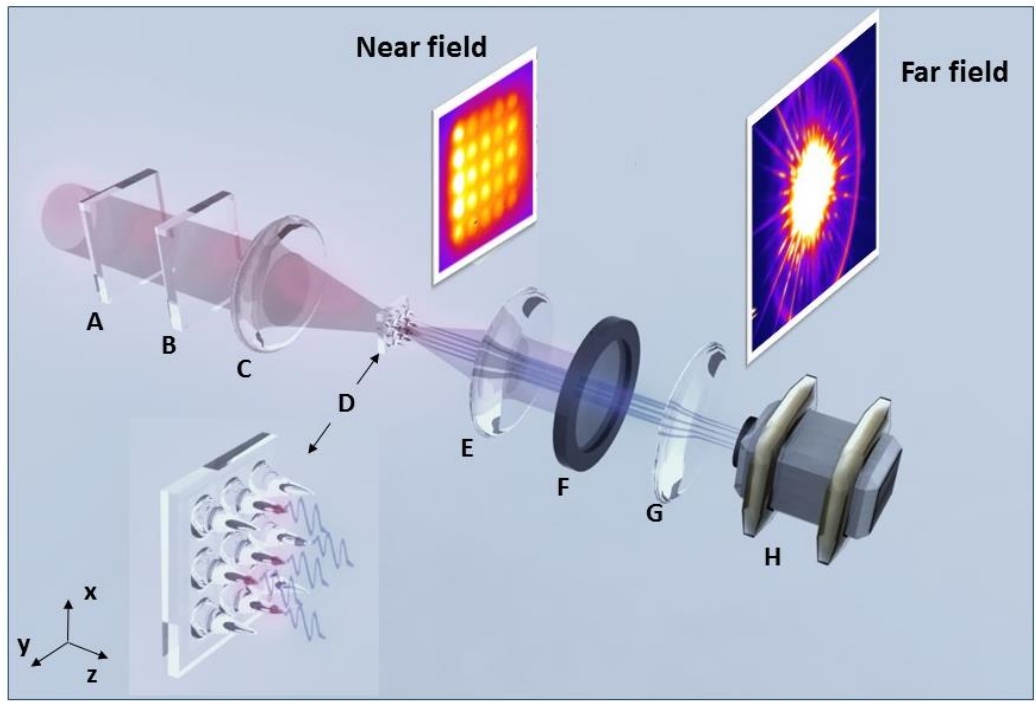

**Figure 2 | Experimental setup.** The mid-infrared laser pump beam is focused onto the nano-structured sample side by a CaF$_2$ plano-convex lens (C) of 20 cm focal length to a focal spot of 80 µm in diameter at FWHM. A set of a mid-infrared half-wave plate (A) and polarizer (B) are used to adjust the laser energy while preserving the laser polarization highly linear along the x axis. The inset is a zoom of the target (D) and sketches the harmonic emission from the nanocones (blue waves). The harmonic radiation emitted from the nanocones is imaged by a set of two lenses (E) and (G) and detected by a CCD-camera (H). Near-field and far-field patterns measured by our optical system have been inserted to illustrate the measurement planes. Far field diffraction patterns illustrate the coherence of the source whereas the near field image show the localization of the amplified harmonic emission. The magnification factor of the nano-emitters plane (near field) to the camera is 40. The 7$^{th}$ harmonic beam is selected by a bandpass filter (F).

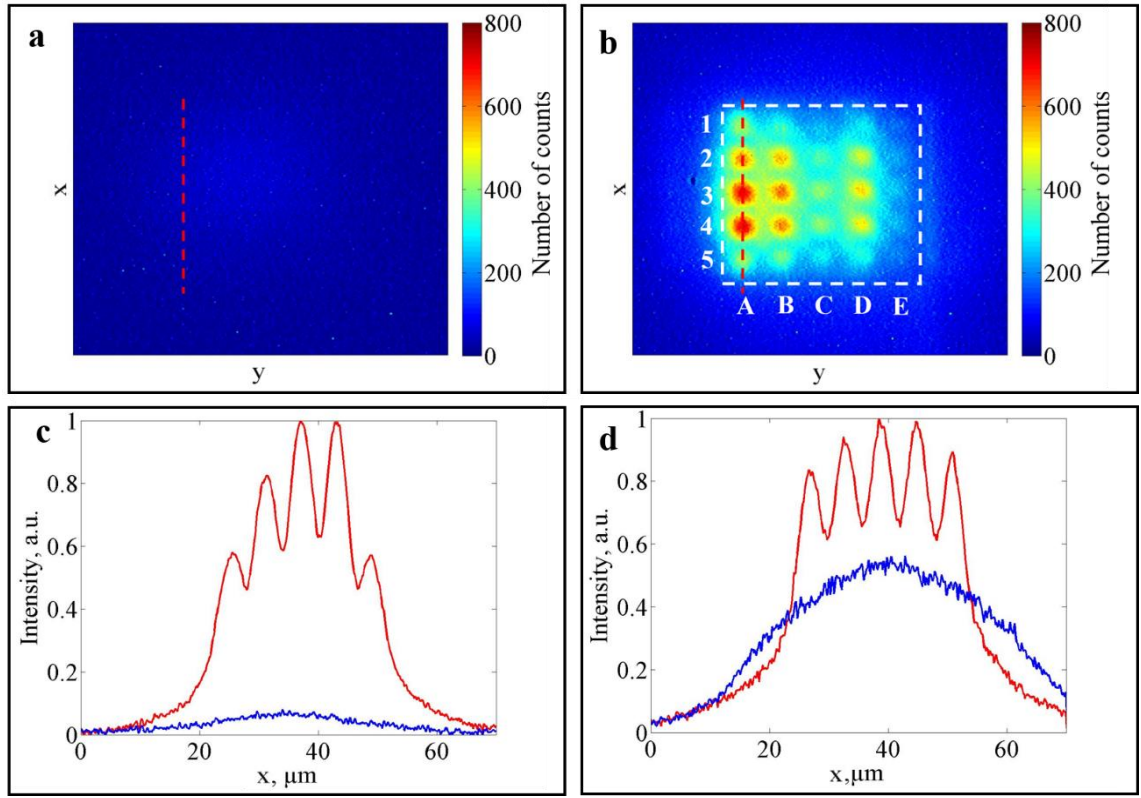

**Figure 3 | Image and intensity profile of the** $7^{th}$ **harmonic beam. a,** $7^{th}$ harmonic signal emitted from the bare ZnO crystal at an intensity of 0.35 TW/cm$^2$. **b,** $7^{th}$ harmonic signal emitted from the ZnO nano-emitters at an intensity of 0.35 TW/cm$^2$. For comparison, the color scale is the same as the one used for Figure 3a. The white rectangle indicates the edges of the array of nano-emitters. Columns are indicated as letters similarly as Figure 1a. **c,** Intensity profile along columns A (red dashed line in Figure 3b) of the $7^{th}$ harmonic signal emitted from ZnO nano-emitters (red) and along the same axis (red dashed line in Figure 3a) from bare crystal (blue) at an intensity of 0.35 TW/cm$^2$. The absolute normalization respects the color scales indicates in Figures 3a and 3b. **d,** Intensity profile along columns A of $7^{th}$ harmonic signal emitted from ZnO nano-emitters (red) and along the same axis from bare crystal (blue) at an intensity of 0.8 TW/cm$^2$ (see supplementary information).

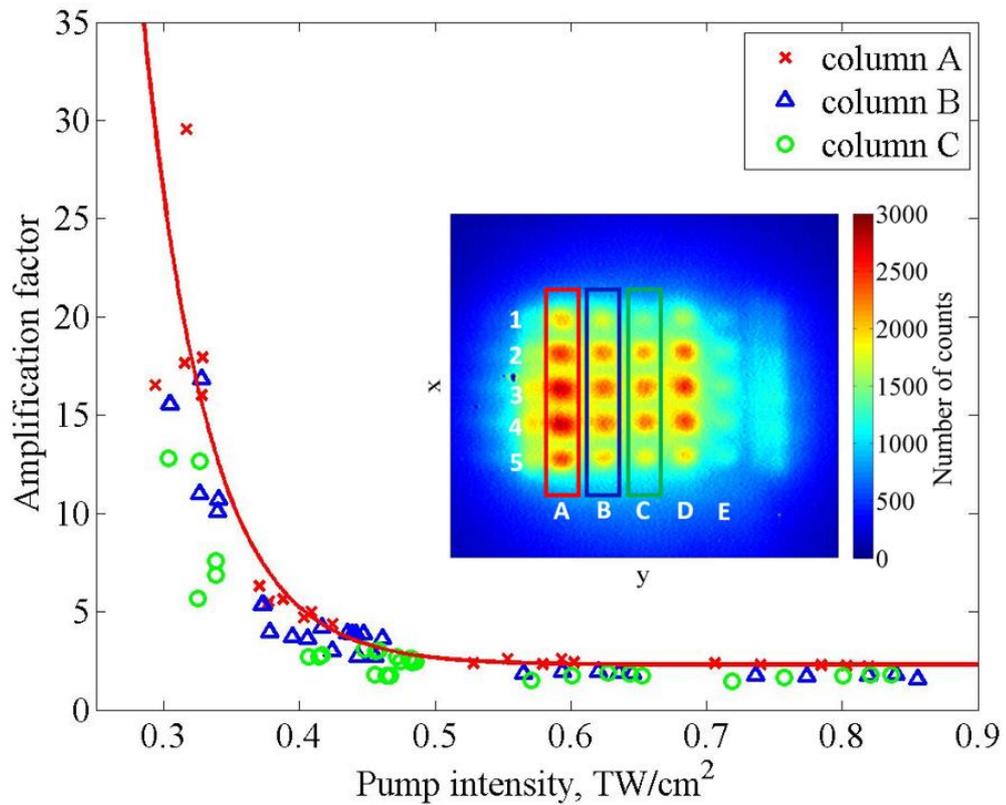

**Figure 4 | Amplification factor of the 7$^{th}$ harmonic in dependence on the pump intensity.** Amplification factors for different columns of nano-emitters in dependence of the intensity of the laser pump beam. The amplification factors of nanocones from columns A, B and C as a function of the laser intensity are shown in Figure 4 as red crosses, blue triangles and green circles, respectively. **Inset:** 7$^{th}$ harmonic image from the ZnO nano-emitters measured at an intensity of 0.6 TW/cm$^2$.

## Methods

**FDTD simulation of the field enhancement**

Finite difference time domain (FDTD) calculations using LUMERICAL Solutions have been performed to optimize the local enhancement of the electromagnetic field in the nanocones at a wavelength of 3.2 µm. The nanocones were constructed according to the shape function $\cos(ax)^4 \times \sin(ay)^4$, where the parameter '$a$' was chosen such that the base size of the nanocones is 4µm. The material of both, the nanocones and their substrate, is ZnO. Its optical parameters were imported from Bond et al., 1965 to LUMERICAL and used throughout the simulations. The surrounding medium is air. The structures are illuminated by a monochromatic plane wave at 3.2 µm wavelength from below the substrate. We used periodic boundary conditions of 6 µm periodicity on both x and y to better represent the experimental case, while perfectly matching layers (PML) were used on the z axis. A finite mesh of 25 nm was used on x, y, and z. The calculation time is about one hour. The supplementary movie M2 was simulated under the same conditions (same structure dimensions and numerical parameters) for wavelengths from 0.8 µm to 4.0 µm in steps of 0.2 µm. The calculation time was 19 hours on a bi-processors Intel Haswell 10C E5-2650V3 (10 cores, 20 threads, max. frequency 3 GHz, Bus speed 9.6 GT/s QPI, 768 GB registered SDRAM (DDR4 2133, 68GB/s bandpass).

The 5 supplementary movies M3, M4, M5, M6 and M7 are simulations of an isolated nanocone base size of 1.6 µm, 2.2 µm, 3.2 µm, 4 µm and 5 µm respectively. We used perfectly matching layers at the boundaries (x-, y- and z-axis) and a finite mesh of 35 nm. The calculation time was between four and up to seventeen hours per movie for larger structures.

**Experimental setup**

As shown in Fig. 2, the pump beam was attenuated by means of a half-wave plate (A) and a polarizer (B) to energies of only a few micro-joules and focused (C) into the sample (D) on the nano-structured sample side. The sample is fixed on a 3D nano-positioning system (Smaract). The harmonic radiation emitted from the nanocones is collimated (E). The seventh harmonic is selected by a a filter (F) and focused (G) onto the CCD camera (H). The integration time of the CCD camera was in the millisecond range and was adjusted to avoid saturation of the detector.

**A**: half-waveplate (Edmund Optics, 85117)
**B**: polarizer (Thorlabs, WP25H-B)
**C**: pcx lens, f = 200 mm, $CaF_2$, E-coating (Thorlabs, LA5714-E)
**D**: ZnO crystal, [11-20], 5 x 5 x 0.5 mm, 2sp (MTI Corporation)
**E**: pcx lens, f = 25 mm, BK7, uncoated (Thorlabs, LA1951)
**F**: 84782 filter (Edmund optics)
**G**: pcx lens, f = 1000 mm, $SiO_2$, uncoated (Thorlabs, LA4184)
**H**: CCD-camera (Imagingsource, 21BUC03)

**The laser system**

The clock for the system is a two-color Erbium-doped fiber laser (Toptica Photonics AG) which provides optically synchronized pulses at 1550 nm and 1050 nm. The Nd:Vanadate pump laser operates at 160 kHz and is electronically synchronized to the fiber laser with a jitter level below 300 fs. Mid-IR pulses at 3200 nm are directly generated from the fiber laser output via difference frequency generation (DFG). The mid-IR seed is amplified in 3 pre-amplifier stages, chirp inverted and then boosted further in 4 power amplifier stages. Finally, compression is achieved though dispersive propagation in 20 cm of Sapphire with a final output of 118 µJ and a duration of 97 fs (sub-9 optical cycles). The output corresponds to an average power of 21 W at 3.2 µm with a pulse-to-pulse output energy stability of 0.33% rms over 30 min.

**Sample production**

The sample were milled using a focused ions beam (FIB) operated by the CSNSM Laboratory in Orsay. The equipment consists of a LEO 1530 SEM (Field Effect Gun) equipped with an Orsay Physics column. The energy of the ion beam (Ga +) is set at 40 keV. We used a high ion beam current of 2 nA, leading to a spot size of several 100s of nm. The total time to produce one nanocone array was 20 minutes.

# Supplementary Information for

## Amplification of high harmonics in 3D semiconductor waveguides

Dominik Franz[1], Rana Nicolas[1], Willem Boutu[1], Liping Shi[2], Quentin Ripault[1], Maria Kholodtsova[1], Bianca Iwan[1], Ugaitz Elu Etxano[3], Milutin Kovacev[2], Jens Biegert[3,4] and Hamed Merdji[1*]

[1] *LIDYL, CEA, CNRS, Université Paris-Saclay,*
*CEA Saclay 91191 Gif sur Yvette France*

[2] *Leibniz Universität Hannover, Institut für Quantenoptik,*
*Welfengarten 1, D-30167 Hannover, Germany*

[3] *ICFO – The Institute of Photonic Sciences,*
*Mediterranean Technology Park, Av. Carl Friedrich Gauss 3, 08860 Castelldefels, Spain*

[4]*ICREA, Pg. Lluís Companys 23, 08010 Barcelona, Spain*

[*]*correspondence to: hamed.merdji@cea.fr*

1) **FDTD simulations and movies M1, M3 -7 details**

The supplementary movie M1 (see supplementary Movie file M1) shows FDTD simulation of the intensity enhancement for different driving laser wavelengths (from 2 µm to 4 µm in steps of 100 nm) in a nanocone with a basis size of 4 µm. Boundary conditions are set for a periodic arrangement of the nanocones fitting with our experimental geometry. While short laser wavelength at 2 µm are confined close to the tip of the nanocone which yields to a high intensity enhancement, longer wavelength at 4 µm hardly penetrate into the cone, resulting in a weak enhancement. Shorter wavelengths exhibit a high enhancement but with an inhomogeneous amplification mode.

At our operating wavelength of 3.2 µm, the maximum intensity enhancement located around 3 µm above the baseline of the nanocones reaches 21 with a nice and regular mode. The parameters have been chosen in order to obtain a smooth field enhancement, favorable for high harmonic local phase matching. Moreover, the control of the laser phase gradient should also favor tailoring the spatial and spectral phases of the harmonic emission, mandatory for the emission of attosecond pulses.

The supplementary movies M3 to M7 (see supplementary Movies M3 to M7) shows FDTD simulations cases that extend our proposal to single nano-emitters. Movies M3 to M7 are the wavelength optimization for nanocone base size of 1.6 µm, 2.2 µm, 3.2 µm, 4 µm and 5 µm respectively. A wavelength step of 200 nm has been taken for each movie. PML boundary conditions are set to respect the nanocone isolation. Note that in this case, the collective field enhancement contribution (around a factor 2) is lost. In all cases, short wavelengths couple well into the nanocone. A nice mode is obtained when the wavelength gets close or longer than the nanocone base size. When the wavelength gets longer, the field enhancement tends to vanish until the electric field does not propagate anymore.

2) **Image and movie showing the amplification of the 7$^{th}$ harmonic signal at different intensities from the bar crystal and the nanocones.**

To explore the amplification regime, we have varied the intensity of the pump beam in the range 0.30 – 0.90 TW/cm$^2$ (see Figure S1). While at a pump intensity of 0.35 TW/cm$^2$ there is almost no 7$^{th}$ harmonic signal from the bare crystal (Figure S1 a), the signal from the nanocones is strong (Figure S1 b). This is different at higher pump intensity around 0.80 TW/cm$^2$ where a strong 7$^{th}$ harmonic is generated from the bare crystal nanostructures (Figure S1 c). At this intensity, the amplification of the 7$^{th}$ harmonic in the nanocones saturates (Figure S1 d) and is only slightly stronger than the one from the bare crystal All images in Figure S1 are normalized to an integration time of 1 second.

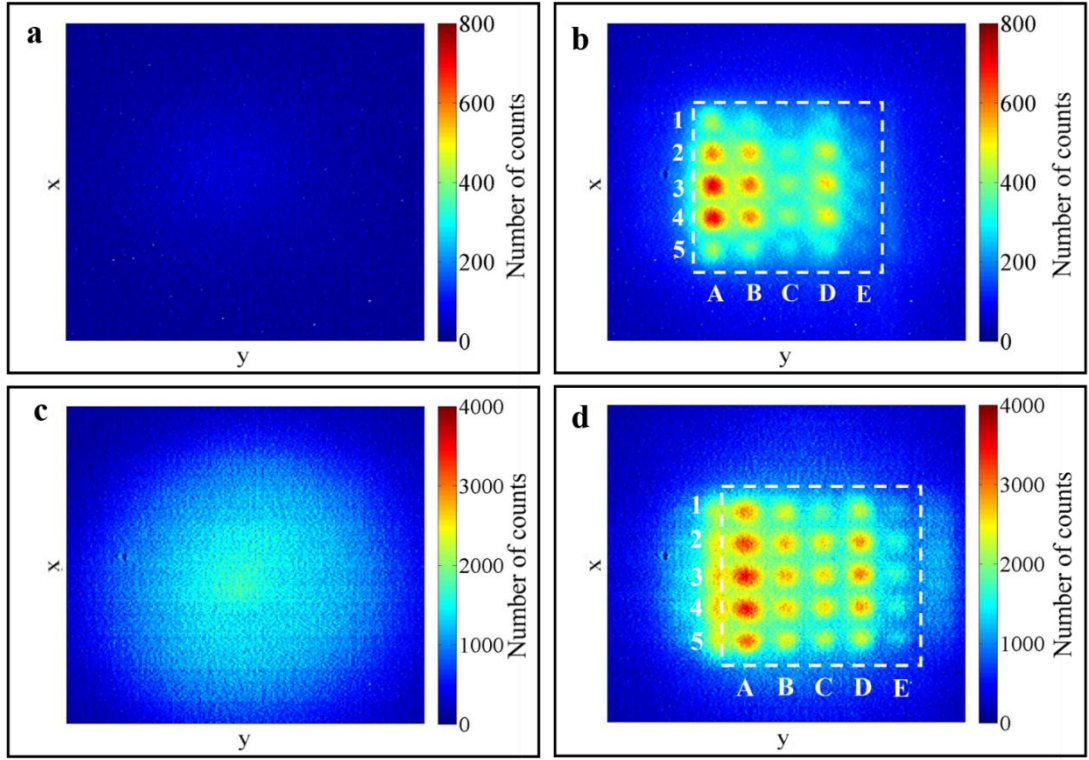

**Supplementary Figure | S1. H7 from nanocones and bare crystal for 0.35 and 0.80 TW/cm$^2$. a,** 7$^{th}$ harmonic from bare crystal at 0.35 TW/cm$^2$ **b,** 7$^{th}$ harmonic from nanocones at 0.35 TW/cm$^2$ **c,** 7$^{th}$ harmonic from bare crystal at 0.80 TW/cm$^2$ **d,** 7$^{th}$ harmonic from nanocones at 0.80 TW/cm$^2$

The dynamic amplification process is show in movie M2 (see supplementary movie file M2). Data were taken in the strong amplification regime at an intensity of 0.65 TW/cm$^2$. The movie has been realized by assembling a transverse scan of the nanocones across the Gaussian beam at an integration time of 200 ms. The 7$^{th}$ harmonic signal is monitored by using the same imaging setup show in the experimental setup (Fig. 2). The first sequence of the movie shows the 7$^{th}$ harmonic signal from the bare crystal. The nano-positioning system then moves the sample step by step (3 µm steps) and the CCD acquires images at each position, with the sample out of the laser beam focus at the beginning of the movie, centered in the focus of the beam at the middle of the movie and out of the focus at the end. All the images are then assembled and combined to a mp4-movie.

3) **Analysis of the nanocones fabrication structuration**

The patterning process using the FIB is applied by scanning the gallium ion beam row by row. Consequently, ZnO can be re-deposited from one row to the other. In order to analyze the amount of re-deposition that occurs during the FIB patterning process, one half of nanocones 5A and 5C was cut and removed by FIB milling, leaving their symmetrical counterparts. The crystalline part and the redeposited part can be distinguished in the SEM images shown in Figure S2. The analysis of the cross-section of the nanocones reveals that nanocone 5A (Figure S2 a), that was patterned last, has much less re-deposition (the redeposited matter appears in dark green, it is up to 200 nm thick) than nanocone 5C (Figure S2 b, the redeposited matter appears in bright green, it is up to 1 µm thick). The re-deposited matter has the same composition as the crystal as shown by energy dispersive X-ray (EDX) measurements localized on several area of the sample as shown in Figure S3. Note that the gallium ions used during the milling process is found in negligible quantity in the sample.

Figure S2b image exhibits clearly different contrasts between the pure crystal ZnO and the re-deposited ZnO. This is a signature of different phases. The re-deposited ZnO should be amorphous and of lower density. This may then strongly affect the coupling of the pump-beam into the nanocones as well as the out-coupling of the 7$^{th}$ harmonic into air and explains the lower amplification efficiency of row C compared to row A. Further optimization of the fabrication techniques with the goal of minimizing the amount of re-deposited matter should easily result in an increase of the light coupling.

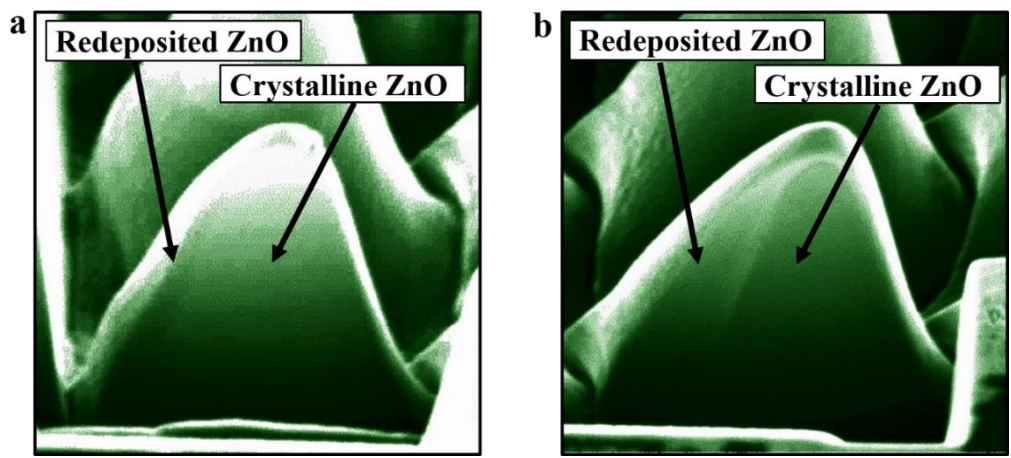

**Supplementary Figure | S2. Analysis of the amount of re-deposited matter during FIB. a,** Nanocones in column 5A have a thin layer of re-deposited matter (dark green) with a thickness of up

to 300 nm. **b,** Nanocones in column 5C have a comparably thick layer of re-deposited matter (here the layer appears brighter than the crystalline part) with a thickness of up to 1 µm. The picture shows that the crystalline part of the nanocones is symmetric and follows the $\cos^4$-behaviour according to which they were designed.

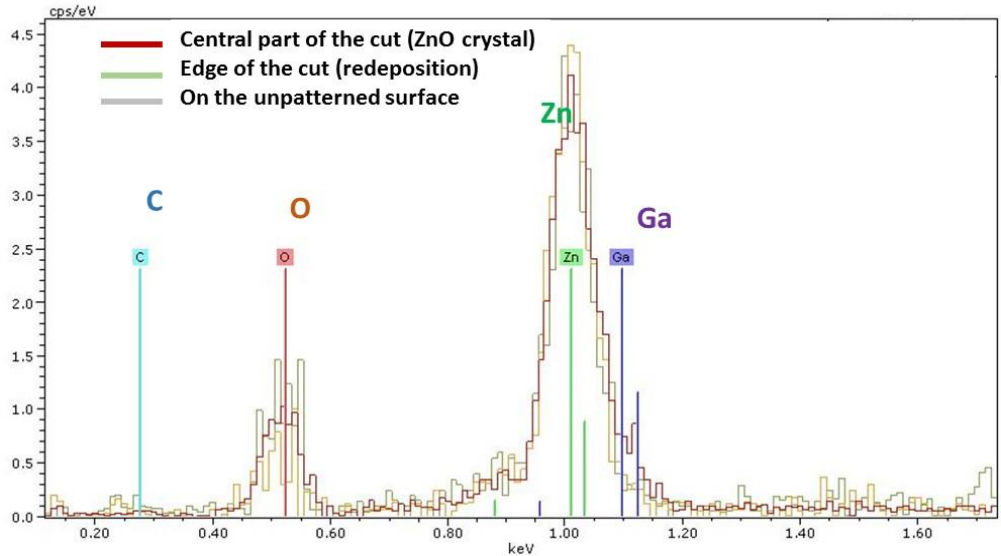

**Supplementary Figure | S3.** Energy dispersive X-ray analysis of the bare crystal, crystal nanocones and re-deposited material. The spatial resolution of the EDX analyser is better about 1µm.

4) **Damage of the nanocones**

While the global shape of the nanocones is unaffected by laser irradiation, a black spot appears at the tip of most of the nanocones (Figure S4 a). Despite these melting features at the tip of the nanocones, remarkably, no change in harmonic signal could be observed during long exposure times. In Figure S4b a zoom at the tip of another nanocones array irradiated at high intensity shows a peeling of the redeposited matter which thickness is in agreement with the observed layer shown in Figure S2. The fact that melting features are only observed on the nanocones and not on the bare crystal is attributed to the high field enhancement in the nanocones.

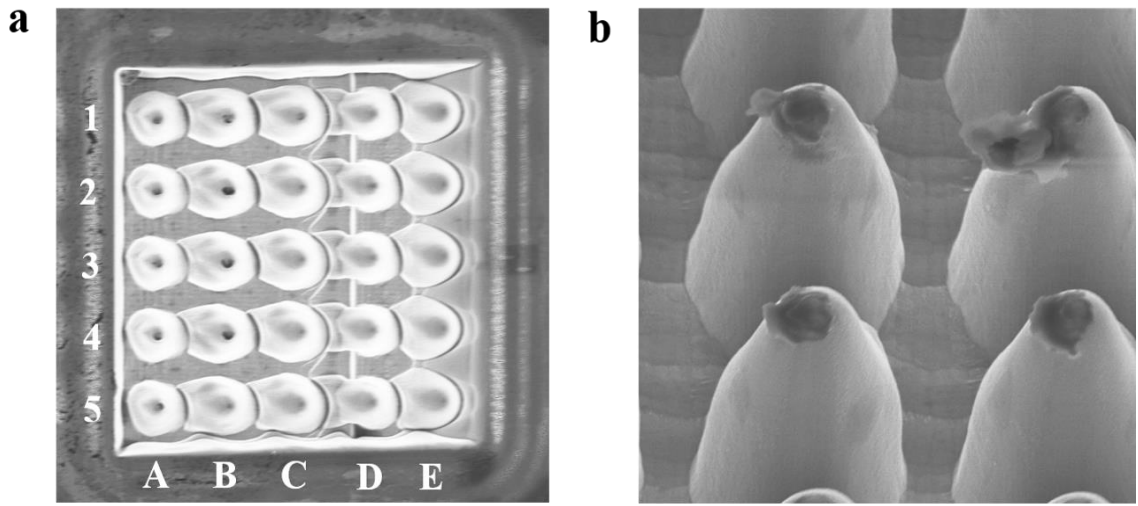

**Supplementary Figure | S4 Nanocones after irradiation. a,** Nanocones view at an angle of 0°. Nanocones with little re-depositions of matter have melting features at the tip (mostly column A, but even column B), which is not the case for nanotips with re-deposition **b,** Zoom on the tip of nanocones at view at an angle of 45°. Ablation of matter at the tip of the nanocones can be observed.

### 5) Far field coherent diffraction patterns

The optical system (see Figure 2) has been set to image the far field diffraction patterns produced by the emission of the array of nanocones. This is illustrated by the far field profiles shown in Figure S5. Due to the size of the individual nano-sources, the divergence of the beam is extremely high (0.5 rad for the 7$^{th}$ harmonic). The images displayed in Figure S5.a and S5.b are recorded by imaging a far field plane located after the sample. In Figure S5.a no filter has been used so that all the harmonics are collected by the detector. In this case, the comb of harmonics leads to a varying period of the interferometric peaks. In Figure S5.b a filter selects only the 7$^{th}$ harmonic and regular diffraction peaks are observed. The actual data were taken with a low dynamic CCD camera which does not allow applying a phase retrieval algorithm to reconstruct the images of the nanocones. Note that the near field imaging spatial resolution was enough to obtain the localization of the array of harmonic nano-sources. However, for much shorter wavelength, typically below 100 nm, phase retrieval approaches can lead to sub-100 nm spatial localization of the harmonic sources. The spatial phase can then be used to get insight into the underlying field confinement effect and single atom effects in semiconductor devices.

To extend this case to shorter wavelengths, nanocones emission has been simulated and is shown in Figure S6 for a single harmonic (13$^{th}$) of an array of nanocones excited at 800 nm. Note that similar harmonics have already been experimentally reported by Han et al.[1]. The observation of interferometric modulation gives a clear signature of the coherence of the generation process. The zoom inside the Airy disk shown in Figure 6c is qualitatively similar to the one measured in our experiment and shown in Figure S5b. At this harmonic wavelength ($\lambda = 61$ nm) it is difficult to get a direct image of the XUV emission. However, by using a coherent diffractive imaging or a holographic scheme we can reconstruct the complete spatial amplitude and phase distribution of the harmonic XUV nano-sources [2,3]. This can be applied to reveal the localization and the strengths of the enhanced electric field for near field phase matching.

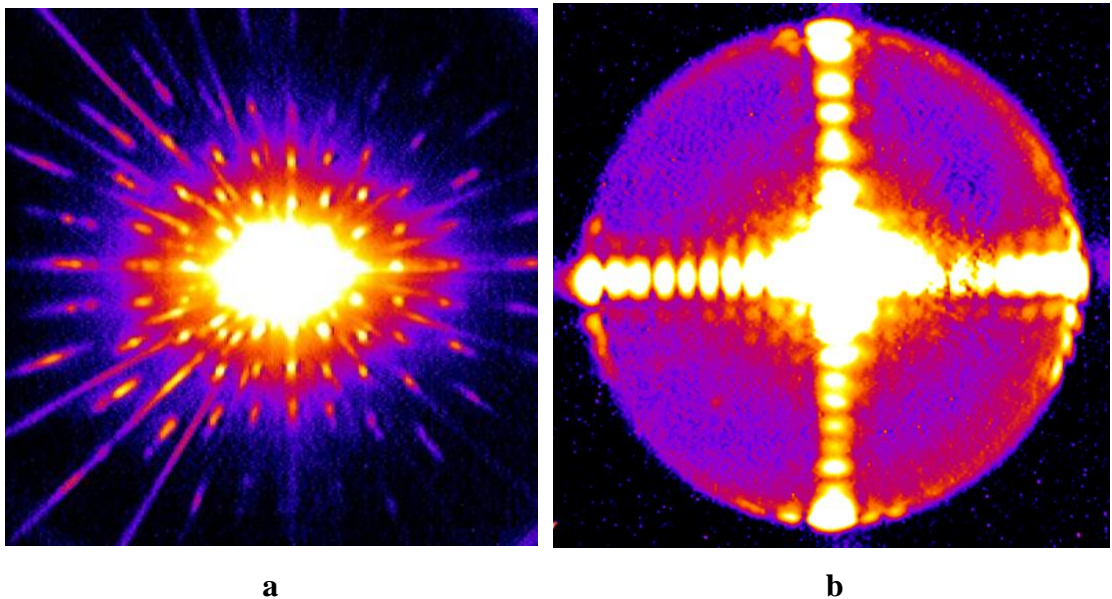

a                                  b

**Supplementary Figure | S5:** Measurement of the far field harmonic emission far harmonic emission recorded : (a) close to the nanocone array for all harmonics; (b) far from the nanocone array for the 7$^{th}$ harmonic.

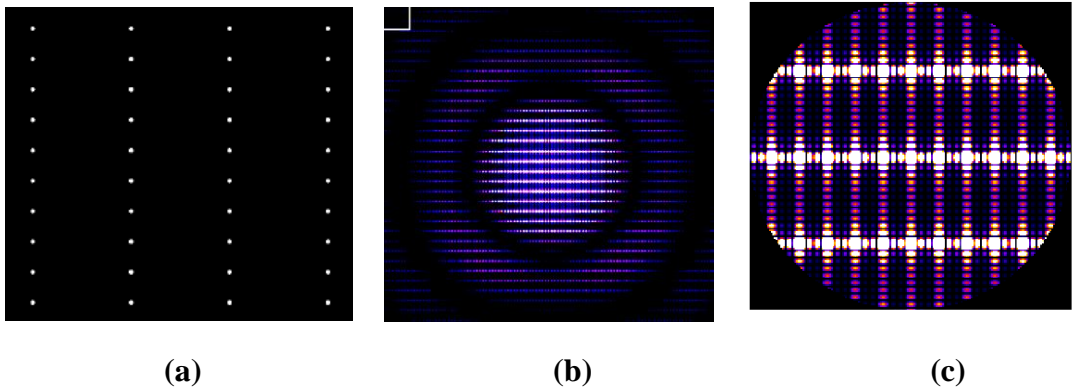

(a)            (b)            (c)

**Supplementary Figure | S6:** (a) Simulation distribution of our array of nanocones HHG sources. (b) Far field emission of the HHG nanocones array shown in (a). (c) zoom of the first Airy disk which show modulations from the interefences of the mutually coherent HHG sources.